\documentclass[twocolumn,prb,aps,nobalancelastpage,amssymb,citeautoscript,hypertex]{revtex4}

\usepackage{graphicx}
\usepackage{hyperref}

\begin{document}

\title{Superconductor-Insulator Phase Separation Induced by \\Rapid Cooling in $\kappa-$(ET)$_2$Cu[N(CN)$_2$]Br}

\author{O.J. Taylor,$^1$ A. Carrington,$^1$ and J.A. Schlueter,$^2$}

\affiliation{$^1$H. H. Wills Physics Laboratory, University of Bristol, Tyndall Avenue, BS8 1TL, United Kingdom}

\affiliation{$^2$Materials Science Division, Argonne National Laboratory, Argonne, Illinois 60439, U.S.A.}

\date{\today}
\begin{abstract}
We present measurements of the low temperature specific heat of single crystals of $\kappa-$(ET)$_2$Cu[N(CN)$_2$]Br as
a function of the cooling rate through the glasslike structure transition at $\sim$ 80K. We find that rapid cooling
produces a small ($\lesssim 4$\%) decrease in the superconducting transition temperature accompanied by a substantial
(up to 50\%) decrease in the normal-state electronic specific heat.  A natural explanation of our data is that there is
a macroscopic phase separation between superconducting and insulating regions in rapidly cooled samples.
\end{abstract}

\pacs{}%
\maketitle

Organic superconductors based on the ET [bis(ethylenedithio)-tetrathiafulvalene] molecule, with general formula
(ET)$_2$X, are composed of conducting cation (ET) layers separated by `insulating' anion (X) layers. The weak overlap
between the conducting layers means that their electronic properties are quasi-two-dimensional. These materials display
a rich phase diagram as a function of temperature and pressure.  For example,  at low temperature and ambient pressure
$\kappa-$(ET)$_2$Cu[N(CN)$_2$]Cl is an antiferromagnetic insulator (AFI).  Application of moderate pressure ($\sim$300
bar) causes an insulator-superconductor transition (IS) with a maximum $T_c\simeq $13K \cite{LefebvreWBBJMFB00}. Close
to this transition there is multiphase region where the superconducting and insulating phases coexist
\cite{LefebvreWBBJMFB00,LimeletteWFGCPJMB03}. There is strong evidence that the superconductivity is unconventional,
with $d$-wave like nodes in the superconducting energy gap \cite{SingletonM02,TaylorCS07}. Although the phase diagram
is similar to the high $T_c$ cuprate superconductors, here the pressure induced IS transition is caused by a reduction
of the ratio of on-site Coulomb repulsion $U$ to the conduction electron bandwidth $W$, rather than a change in carrier
density \cite{Kanoda97,Mckenzie97,SingletonM02}. The position at ambient pressure of different compounds in the series
is controlled by the anion and/or the degree of deuteration of the ET molecules, both of which can be thought of as
applying `chemical pressure'.

A widely studied member of the $\kappa$ phase materials is $\kappa-$(ET)$_2$Cu[N(CN)$_2$]Br (hereafter abbreviated to
$\kappa$-Br), which is a superconductor with a $T_c$ of $\sim$ 12.4~K. At ambient pressure $\kappa$-Br sits close to
the AFI phase boundary, and deuteration causes it to move even closer to this boundary \cite{TaniguchiKK03}. At
$T_g\simeq 77$K there is a glasslike structural transition \cite{MullerLSSKS02} and the cooling rate through this
temperature strongly effects the normal and superconducting state properties. The nature of this structural transition
is unclear. A widely held hypothesis is that it is associated with a configuration change in the order of the terminal
ethylene groups of the ET molecules \cite{MullerLSSKS02}.  Although this theory is supported by the existence of an
isotope effect on $T_g$, a recent high resolution x-ray structural study\cite{WolterFDSSLS07} found that, even in fast
cooled samples, the ethylene groups are almost completely ordered at the lowest temperatures (9K) . It was suggested
 that the disorder may instead be associated with the polymeric anion chain.\cite{WolterFDSSLS07}

One consequence of rapid cooling through $T_g$ in this compound is a reduction in the superconducting transition
temperature $T_c$\cite{TokumotoKTKA99,SuZSKW98}. This effect has been shown to vary over four orders of magnitude of
cooling rate\cite{TokumotoKTKA99}.  Magnetization measurements have shown that fast cooling also causes a marked
decrease in the magnetic screening which was interpreted as either a decrease in the superconducting volume fraction or
an increase in the magnetic penetration depth $\lambda$ \cite{AburtoFP98}. Scanning microregion infrared reflectance
spectroscopy (SMIRS) measurements\cite{YoneyamaSKIK05} have shown evidence for macroscopic insulating/metallic region
phase separation at the \textit{surface} of fast cooled samples. In deuterated $\kappa$-Br $^{13}$C-NMR
\cite{MiyagawaKK02} and magnetoresistance measurements\cite{TaniguchiKK03} show evidence for phase separation even in
slowly cooled samples.

In this paper, we report measurements of the specific heat of single crystals of $\kappa$-Br as the cooling rate
through $T_g$ is varied from $\sim$ 0.02 K/min to $\sim$ 100~K/min.  By applying a large magnetic field ($\mu_0H$=14T)
perpendicular to the conducting planes we can suppress the superconductivity and study the evolution of the normal
state electronic specific heat.   We find that that the Sommerfeld constant $\gamma$ is reduced by up to a factor two
by fast cooling which we suggest is caused by insulating/metallic region phase separation occurring throughout the bulk
of the whole sample.

Single crystals of $\kappa$-Br were grown by a standard electrochemical technique \cite{KiniGWCWKVTSJW90}. The specific
heat was measured using a `long relaxation' calorimetry technique \cite{WangPJ01,TaylorCS07} using a bare Cernox
\cite{lakeshore} chip as the sample platform, heater and thermometer.  The performance of the calorimeter was
extensively checked by measuring samples of Ag. The maximum absolute error was $\sim \pm 1$\%. The field dependence of
the Cernox thermometer was measured against a capacitance thermometer and checked by measuring the specific heat of Ag
which is virtually field independent in our temperature range.

Two samples of $\kappa$-Br were measured.  Sample A had a mass of 249~$\mu$g and dimensions $0.66 \times 0.61 \times
0.30$mm$^3$ (the shortest dimension is the low-conductivity $b$ axis) and sample B had a mass of  545~$\mu$g and
dimensions $0.90\times 0.85 \times 0.35$mm$^3$.  These samples were repeatedly cooled down from $T$=85~K to $T$=65~K,
which is the temperature range of the glass transition \cite{MullerLSSKS02}, at cooling rates between 0.02~K/min and
$\sim 100$~K/min, then to $T$=1.3~K at the maximum cooling rate of the cryostat ($\sim$ 1--2~K/min for the slow cooled
samples). Rapid cooling above 1.5~K/min was achieved by admitting $^4$He exchange gas, which was then pumped out while
the sample was held at $\approx20$~K to prevent gas absorption on the calorimeter. The specific heat $C$ was measured
after each cool down at various fields between 0 and 14T, applied perpendicular to the conducting planes. It was shown
previously \cite{TaylorCS07} that in $\kappa$-Br $C$ becomes field independent above $\mu_0H_{c2}\simeq $8 T, and so
$C(\mu_0H=14$T) can be taken as the normal state value.

\begin{figure}
\includegraphics[width=0.8\linewidth,clip]{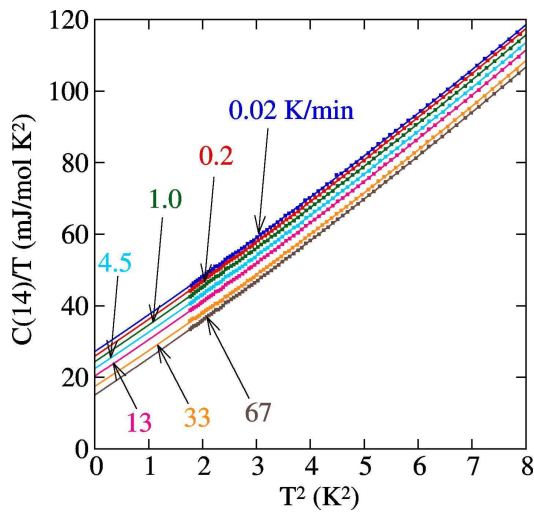}
\caption{(color online). Low temperature normal state specific heat  measured in a field of 14 T for sample A after it
had been cooled, through the glass transition region (85-65K), at the different rates indicated. The lines are second
order polynomial fits.} \label{fig:ct_t^2}
\end{figure}

\begin{figure}
\includegraphics[width=0.8\linewidth,clip]{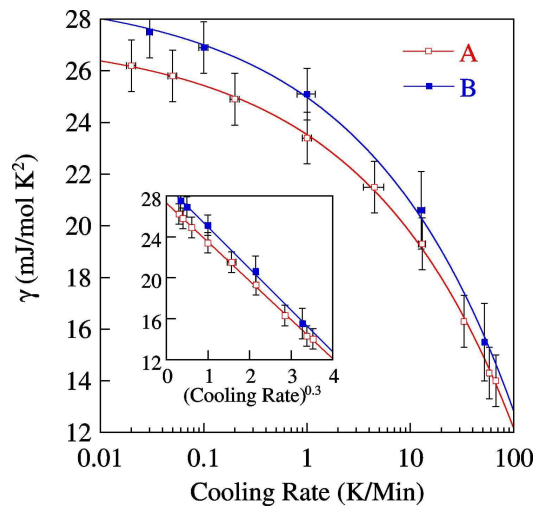}
\caption{(color online).Cooling rate dependence of $\gamma$ for both samples. The inset shows the same data plotted
versus (cooling rate)$^{0.3}$.  The solid lines in both parts of the figure are fits to this power law.}
\label{fig:gamma_rate}
\end{figure}

Fig.~\ref{fig:ct_t^2} shows the 14~T normal state specific heat data plotted as $C/T$ versus $T^2$, for various cooling
rates.  It can be seen directly that there is a significant decrease in the Sommerfeld constant $\gamma$ with
increasing cooling rate.  The data can be fitted by $C/T=\gamma+\beta_3 T^2+\beta_5 T^4$ where where $\beta_3$ and
$\beta_5$ are the coefficients of the leading order phonons terms.  For the slowest cooling rates, $\gamma=26\pm
1$mJmol$^{-1}$K$^{-2}$ and $\beta_3$ corresponds to a Debye temperature of $218\pm10$ K, in agreement with previous
studies \cite{ElsingerWWHSS00}.  As the cooling rate was increased we find that $\beta_3$ and $\beta_5$ remain constant
to within 1\% and 6\% respectively, whereas $\gamma$ decreases by up to almost 50\%.  The dependence of $\gamma$ on
cooling rate is plotted in Fig.~\ref{fig:gamma_rate} for both samples.  Although at the lowest cooling rates there is a
small difference in $\gamma$ between the samples the dependence of $\gamma$ on cooling rate is very similar. As shown
in the inset to Fig.~\ref{fig:gamma_rate} the data approximately obey an empirical power law, $\gamma=\gamma_0 -
\eta(dT/dt)^{n}$, with $n=0.3\pm 0.05$.  This reduction in $\gamma$ is consistent with the reduction in superconducting
volume fraction (or decrease in superfluid density) observed in magnetization measurements. \cite{AburtoFP98}

\begin{figure}
\includegraphics[width=0.8\linewidth,clip]{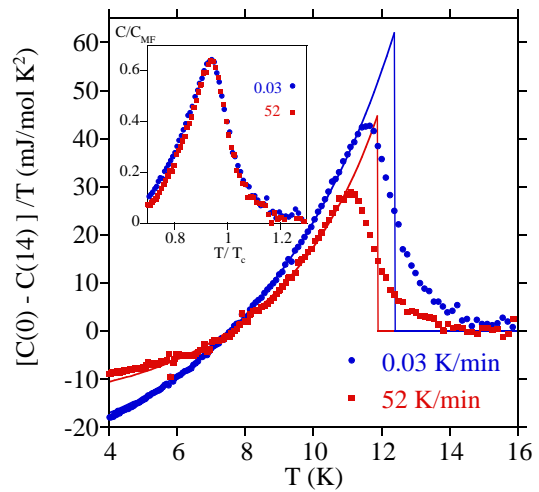}
\caption{(color online). Zero field electronic specific heat [$C(0)-C(14)]/T$ of sample B, for two different cooling
rates.  The solid lines are fits to a mean-field model which is used to determine $T_c$ and the height of the specific
heat jump at $T_c$. Inset: Scaled plot of the data close to $T_c$.} \label{fig:deltaC}
\end{figure}

By subtracting the 14T normal state data from the zero field data the superconducting anomaly at $T_c$ is clearly
discernable (see Fig.~\ref{fig:deltaC}).  The anomaly is a rather small proportion of the total specific heat ($\Delta
C/C \simeq 3$\%).  For simplicity we fit the anomaly to a mean-field theory, neglecting the fact that the anomaly is
broadened both by thermal fluctuations and sample inhomogeneity.  Specifically, we use a strong coupling form of the
mean-field $d$-wave theory which was shown to fit the data from the lowest temperatures right up to $T_c$
\cite{TaylorCS07}.  We note however, that an $s$-wave model works equally well close to $T_c$ \cite{TaylorCS07}.  $T_c$
corresponds to the mid-point of the leading edge of the anomaly, and the extrapolated anomaly height
$\Delta C_{\rm MF}$ is taken from the fit.  The figure shows data for the slowest cooling rate and a fast one. Cooling
at 52K/min causes $T_c$ to decrease by 0.6~K and $\gamma$ to decrease by $\sim$ 40\%.  The inset to this figure
shows the the same data with the axes normalized.  It can be seen that the anomaly does not get significantly broader upon rapid cooling.

\begin{figure}
\includegraphics[width=0.8\linewidth,clip]{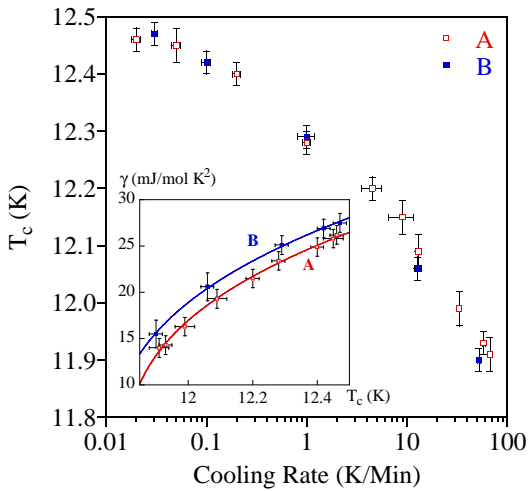}
\caption{(color online).Cooling-rate dependence of $T_c$ for both samples. Inset: $\gamma$ versus $T_c$. The lines are
guides to the eye.} \label{fig:tc_rate}
\end{figure}

The decrease in $T_c$ with increased cooling rate is shown in Fig.\ \ref{fig:tc_rate}. The data for both samples is in
good agreement and also agrees reasonably well with previous studies (see Ref.\ \onlinecite{YoneyamaHSNK04} and
references therein). The $T_c$ reduction at our maximum cooling rate is $\sim$0.6~K or $\sim$ 4\%. We note that here we
have a very small thermometer stage in direct contact with the sample and so the cooling rate registered should be an
accurate reflection of that experienced by the sample. At our slowest cooling rates $T_c$ appears to have saturated at
its maximum value within our resolution ($\pm$~20mK).

\begin{figure}
\includegraphics[width=0.8\linewidth,clip]{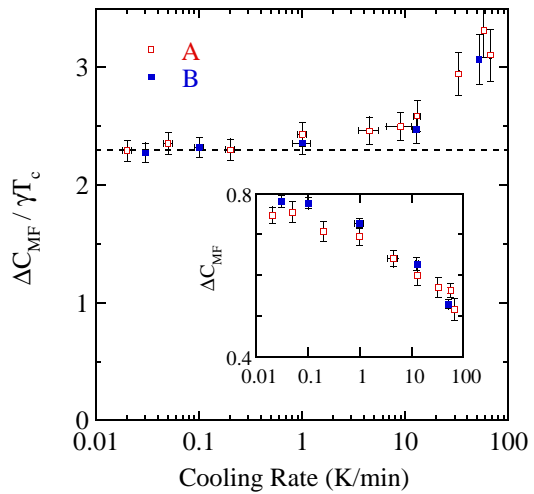}
\caption{(color online). Normalised superconducting anomaly height $\Delta C_{\rm MF}/\gamma T_c$ as a function of
cooling rate for both samples. The inset shows behavior of $\Delta C_{\rm MF}$ in units of Jmol$^{-1}$K$^{-1}$.}
\label{fig:anomaly_height}
\end{figure}

\begin{figure}
\includegraphics[width=0.8\linewidth,clip]{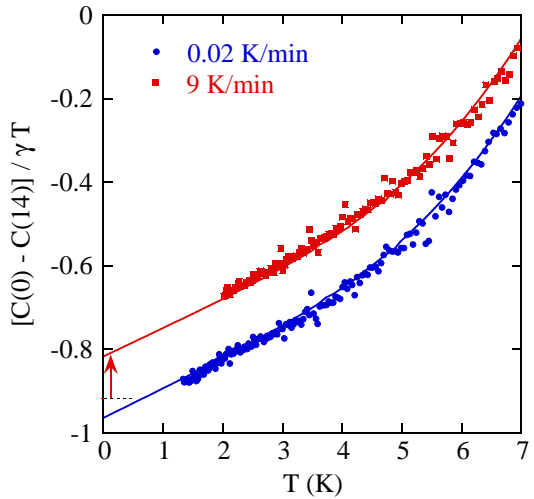}
\caption{(color online). Low temperature behavior of the electronic specific heat in the superconducting state,
$[C(0)-C(14)]/\gamma T$, along with fits to a strong coupling $d$-wave model, at two different cooling rates (sample
A).  The 9K/min data have been offset vertically for clarity by 0.1 as indicated by the arrow.} \label{fig:dwavefits}
\end{figure}

The detailed evolution of the height of the superconducting anomaly can be seen in Fig.\ \ref{fig:anomaly_height}.
Given the dramatic reduction in $\gamma$ the normalized anomaly height is remarkable constant with cooling rate. For
the highest cooling rates $\gtrsim 12$K/min $\Delta C_{\rm MF}/\gamma T_c$ is seen to increase, although this is close
to the resolution limit. For these high cooling rates a small upturn in $C/T$ is seen in the 14T data for $T\lesssim
2$K, probably because of additional magnetic contributions, which increases the uncertainty of our estimates of
$\gamma$.

In Fig.\ \ref{fig:dwavefits} we show the low temperature behavior of the electronic specific heat $[C(0)-C(14)]/\gamma
T$ for sample A at two different cooling rates. In both cases, $[C(0)-C(14)]/\gamma T$ is linear with $T$ below $\sim$
4K, and fits the strong coupling form of the $d$-wave model very well (details of the fit can be found in Ref.\
\onlinecite{TaylorCS07}).  Hence, at least for moderately fast cooling the order parameter symmetry is unaffected.

In conventional models of superconductivity,  the density of states at the Fermi level $N_0$ is an important factor in
determining $T_c$ [in simple BCS theory $T_c\propto \exp(-1/N_0V)$, where $V$ is the superconducting pairing potential
energy]. This continues to be the case even in most more exotic theories, and so it appears very difficult to reconcile
the relatively small decrease in $T_c$ with the large decrease in $\gamma$ (see Fig.\ \ref{fig:tc_rate}) which in
band-theory is proportional to $N_0$.  This behavior is similar to that found upon deuteration of $\kappa$-Br, which
also produces a a large decrease in $\gamma$ with only a small decrease in $T_c$ \cite{NakazawaTKK00}. There is clear
evidence that deuteration moves the system towards the antiferromagnetic state.  In some systems, $gamma$ is found to
diverge at the metal insulator boundary, however the behavior in deuterated $\kappa$-Br is similar to that observed in
the high $T_c$ cuprates.

A natural explanation of our results is that upon fast cooling $\kappa$-Br  phase separates into insulating and
metallic (superconducting) regions.  Given the proximity of $\kappa$-Br to the AFI phase boundary this is plausible and
explains the reduction of the average value of $\gamma$ for the whole sample.  It is also consistent with the SMIRS
results \cite{YoneyamaSKIK05} mentioned above. However, it does not, in itself, explain the observed reduction of
$T_c$. One possibility is that fast cooling causes the structure transition at $T_g$ to be incomplete throughout the
whole sample, and effectively produces a negative pressure moving the system further towards the AFI
phase.\cite{TaniguchiKK03}  As the AFI phase transition is first order, the system naturally may then break up into
superconducting and insulating regions.  This picture also explains the progressive lowering of $\gamma$ as a function
of increased deuteration.\cite{NakazawaTKK00}  The small reduction in $T_c$ could then result, at least partially, from
small changes in the intrinsic density of states and/or pairing interaction strength as the phase diagram is
transversed. This is similar to the behavior observed upon deuteration of $\kappa$-Br, where $T_c$ at first rises
slightly and then falls as the AFI boundary is approached.\cite{YoneyamaSK04} However, in $\kappa$-Cl the opposite
trend is found with $T_c$ being maximum close to the phase boundary \cite{WilliamsKWCGMPWKBCKSOJW90}). Another factor
which needs to be taken into account is the direct effect of disorder. As $\kappa$-Br has a strongly anisotropic energy
gap, even non-magnetic impurities are expected to decrease $T_c$ rapidly, as observed experimentally for $\kappa$-NCS
\cite{AnalytisABOGJP06}. The correlation of the increase in residual resistance $\rho_0$ with the decrease of $T_c$ as
the cooling rate is increased in $\kappa$-Br has been found to be in agreement with that expected for a $d$-wave
superconducting energy gap \cite{PowellM04}. We note however, the presence of insulating regions in rapidly cooled
samples, as suggested by the present work, would also cause a substantial increase $\rho_0$ in addition to the direct
effect of disorder. Hence $\rho_0$ may overestimate the true level of disorder present in the superconducting fraction
of the fast cooled samples.

In summary, we have measured the low-temperature specific heat $\kappa$-Br as a function of cooling rate through the
structural phase transition at $T_g$ ($\simeq 78$~K).  $T_c$ decreases with increased cooling rate, and is accompanied
by a sharp decrease in the normal state electronic specific heat.  This reduction is up to $\sim 50~\%$ at our maximum
cooling rate ( $\sim 100$~K/min). We suggest that this reduction in $\gamma$ is due to phase separation of
superconducting (metallic) and insulating regions, caused by the fast cooling effectively applying negative pressure to
the material and thus driving it closer to the first order antiferromagnetic insulating state.

We thank A.\ Bangura, R.\ Giannetta, and B.\ Powell for helpful discussions. This work was supported by the EPSRC (UK).
Work at Argonne National Laboratory is sponsored by the U.\ S.\ Department of Energy, Office of Basic Energy Sciences,
Division of Materials Sciences, under Contract DE-AC02-06CH11357.

\end{document}